\documentclass[aps,prd,twocolumn,nofootinbib,showpacs]{revtex4}

\usepackage{graphicx,color}
\usepackage[colorlinks=true, linkcolor=blue, citecolor=blue, urlcolor=blue]{hyperref}
\usepackage[cp1251]{inputenc}

\begin{document}

\title{ \begin{center}  Thrust Distribution for 3-Jet Production from $e^+e^-$ Annihilation \\ within the QCD Conformal
Window and in QED\
\end{center} }

\author{Leonardo Di Giustino$^{1}$}
\email[email:]{ldigiustino@uninsubria.it}
\author{Francesco Sannino$^{2,3,4}$ }
\email[email:]{sannino@cp3.sdu.dk}

\author{Sheng-Quan Wang$^{5}$}
\email[email:]{sqwang@cqu.edu.cn}

\author{Xing-Gang Wu$^6$}
\email[email:]{wuxg@cqu.edu.cn}
\address{$^1$Department of Science and High Technology, University of Insubria, via valleggio 11, I-22100, Como, Italy}
\address{$^2$Scuola Superiore Meridionale, Largo S. Marcellino, 10, 80138 Napoli NA, Italy}
\address{$^3$CP3-Origins \& Danish IAS, Univ. of Southern Denmark, Campusvej 55, DK-5230 Odense}
\address{$^4$Dipartimento di Fisica,
E. Pancini, Universita di Napoli Federico II, INFN sezione di Napoli, Complesso Universitario di Monte S. Angelo Edificio 6, via Cintia, 80126 Napoli, Italy
}
\address{$^5$Department of Physics, Guizhou Minzu University, Guiyang 550025, P.R. China}
\address{$^6$Department of Physics, Chongqing University, Chongqing 401331, P.R. China}

\date{\today}

\begin{abstract}

We investigate the theoretical predictions for thrust distribution
in the electron positron annihilation to three-jets process at
NNLO for different values of the number of flavors, $N_f$. To
determine the distribution along the entire renormalization group
flow from the highest energies to zero energy we consider the
number of flavors near the upper boundary of the conformal window.
In this regime of number of flavors the theory develops a
perturbative infrared interacting fixed point. We then  consider
also the QED thrust obtained as the limit $N_c \rightarrow 0$ of
the number of colors. In this case the low energy limit is
governed by an infrared free theory. Using these quantum field
theories limits as theoretical laboratories we arrive at an
interesting comparison between the Conventional Scale Setting -
(CSS) and the Principle of  Maximum Conformality (PMC$_\infty$)
methods. We show that within the perturbative regime of the
conformal window and also out of the conformal window the
PMC$_\infty$ leads to a higher precision, and that reducing the
number of flavors, from the upper boundary to the lower boundary,
through the phase transition the curves given by the PMC$_\infty$
method preserve with continuity the position of the peak, showing
perfect agreement with the experimental data already at NNLO.

\pacs{11.15.Bt,
11.10.Gh,11.10.Jj,12.38.Bx,13.66.De,13.66.Bc,13.66.Jn}

\end{abstract}

\maketitle

\section{Introduction}
We employ, for the first time, the perturbative regime of the
quantum chromodynamics (pQCD) infrared conformal window as a
laboratory to investigate in a controllable manner (near)
conformal properties of physically relevant quantities such as the
thrust distribution in electron positron annihilation processes.

The conformal window of pQCD has a long and noble history
conveniently summarised and generalised to arbitrary
representations in Ref. \cite{Dietrich:2006cm}. This work led to
renew interest in the subject and to a substantial number of
lattice papers whose results and efforts that spanned a decade
have been summarised in a recent report on the subject in Ref.
\cite{Cacciapaglia:2020kgq}.

When all quark masses are set to zero two physical parameters
dictate the dynamic of the theory and these are the number of
flavors $N_f$ and colors $N_c$. Already at the one loop level one
can distinguish two regimes of the theory. For the number of
flavors larger than $11N_c/2$ the theory possesses an infrared
non-interacting fixed point and at low energies the theory is
known as non-abelian quantum electrodynamics (non-abelian QED).
The high energy behavior of the theory is uncertain, it depends on
the number of active flavors and there is the possibility that it
could develop a critical number of flavors above which the theory
reaches an UV fixed point \cite{Antipin:2017ebo} and therefore
becomes safe. When the number of flavors is below $11 N_c /2$ the
non-interacting fixed point becomes UV in nature and then we say
that the theory is asymptotically free.  Lowering the number of
flavors just below the point when asymptotic freedom is restored
the theory develops a trustable infrared interacting fixed point
discovered by Banks and Zaks \cite{bankszaks} at two-loop level.
The analysis at higher loops has been performed in
\cite{Pica:2010xq,Ryttov:2010iz,Ryttov:2016ner}. As the number of
flavors are further dropped it is widely expected that a quantum
phase transition occurs. The nature, the dynamics and the
potential universal behavior of this phase transition is still
unknown \cite{Cacciapaglia:2020kgq}. At lower scales, we
substantially lower the number of matter fields and we observe
chiral symmetry breaking. A dynamical scale is then spontaneously
generated yielding the bulk of all the known hadron masses. The
two-dimensional region in the number of flavors and colors where
asymptotically free QCD develops an IR interacting fixed point is
colloquially known as the {\it conformal window of pQCD}. In this
work we will consider the region of flavors and colors near the
upper bound of the conformal window where the IR fixed point can
be reliably accessed in
perturbation theory.\\
The thrust distribution and the Event Shape variables are a
fundamental tool in order to probe the geometrical structure of a
given process at colliders. Being observables that are exclusive
enough with respect to the final state, they allow for a deeper
geometrical analysis of the process and they are also particularly
suitable for the measurement of the strong coupling $\alpha_s$\cite{Kluth:2006bw}.\\
Given the high precision data collected at LEP and SLAC
\cite{opal,aleph,delphi,l3,sld}, refined calculations are crucial
in order to extract information to the highest possible precision.
Though extensive studies on these observables have been released
during the last decades including higher order corrections from
next-to-leading order (NLO) calculations~\cite{Ellis:1980wv,
Kunszt:1980vt, Vermaseren:1980qz, Fabricius:1981sx, Giele:1991vf,
Catani:1996jh} to the next-to-next-to-leading
order(NNLO)~\cite{Gehrmann-DeRidder:2007nzq,
GehrmannDeRidder:2007hr, Ridder:2014wza, Weinzierl:2008iv,
Weinzierl:2009ms} and including resummation of the large
logarithms~\cite{Abbate:2010xh, Banfi:2014sua}, the theoretical
predictions are still affected by significant theoretical
uncertainties that are related to large renormalization energy
scale ambiguities. In the particular case of the three-jet event
shape distributions the conventional practice (Conventional Scale
Setting - CSS) of fixing the renormalization scale to the
center-of-mass energy $\mu_r=\sqrt{s}$, and of evaluating the
uncertainties by varying the scale within an arbitrary range, e.g.
$\mu_r\in[\sqrt{s}/2,2\sqrt{s}]$ lead to results that do not match
the experimental data and the extracted values of $\alpha_s$
deviate from the world average~\cite{Tanabashi:2018oca}.
Additionally, the CSS procedure is not consistent with the
Gell-Mann-Low scheme~\cite{GellMann:1954fq} in Quantum
Electrodynamics (QED), the pQCD predictions are affected by scheme
dependence and the resulting perturbative QCD series is also
factorially divergent like $n!\beta^n_0\alpha^n_s$, i.e. the
"renormalon" problem~\cite{Beneke:1998ui}. Given the factorial
growth, the hope to suppress  scale uncertainties by including
higher-order corrections is compromised. \\
A solution to the scale ambiguity problem is offered by the
{\emph{Principle of  Maximum Conformality}}
(PMC)~\cite{Brodsky:1982gc,Brodsky:2011ig,Brodsky:2011ta,
Brodsky:2012rj, Mojaza:2012mf, Brodsky:2013vpa}. This method
provides a systematic way to eliminate renormalization
scheme-and-scale ambiguities from first principles by absorbing
the $\beta$ terms that govern the behavior of the running coupling
via the renormalization group equation. Thus, the divergent
renormalon terms cancel, which improves the convergence of the
perturbative QCD series. Furthermore, the resulting PMC
predictions do not depend on the particular scheme used, thereby
preserving the principles of renormalization group
invariance~\cite{Brodsky:2012ms, Wu:2014iba}. The PMC procedure is
also consistent with the standard Gell-Mann-Low method in the
Abelian limit, $N_c\rightarrow0$~\cite{Brodsky:1997jk}. Besides,
in a theory of unification of all forces, electromagnetic, weak
and strong interactions, such as the Standard Model, or Grand
Unification theories, one cannot simply apply a different
scale-setting or analytic procedure to different sectors of the
theory. The PMC offers the possibility to apply the same method in
all sectors of a theory, starting from first principles,
eliminating the renormalon growth, the scheme dependence, the
scale ambiguity, and satisfying the QED Gell-Mann-Low scheme
in the zero-color limit $N_c\to 0$.\\
In particular, recent applications of the PMC and of the
{\emph{Infinite-Order Scale-Setting using the Principle of Maximum
Conformality}} (PMC$_\infty$) have shown to significantly reduce
the theoretical errors in Event Shape Variable distributions
highly improving also the fit with the experimental
data\cite{DiGiustino:2020fbk} and to improve the theoretical
prediction on $\alpha_s$
with respect to the world average \cite{Wang:2019ljl}\cite{Wang:2019isi}.\\
It would be highly desirable to compare the PMC and CSS methods
along the entire renormalization group flow from the highest
energies down to zero energy. This is precluded in standard QCD
with a number of active flavors less than six because the theory
becomes strongly coupled at low energies. We therefore employ the
perturbative regime of the conformal window which allows us to
arrive at arbitrary low energies and obtain the corresponding
results for the SU($3$) case at the cost of increasing the number
of active flavors. Here we are able to deduce the full solution at
NNLO in the strong coupling. We consider also the U(1) abelian QED
thrust distribution which rather than being infrared interacting
is infrared free. We conclude by presenting the comparison between
two renormalization scale methods, the CSS and the PMC$_\infty$.


\subsection{The Strong Coupling at NNLO}

The value of the QCD strong coupling $\alpha_s$ at different
energies $\mu$ can be computed via its $\beta$-function:
\begin{equation}
\mu^{2} \frac{d}{d \mu^{2}}\left(\frac{\alpha_{s}}{2
\pi}\right)=-\frac{1}{2} \beta_{0}\left(\frac{\alpha_{s}}{2
\pi}\right)^{2}-\frac{1}{4} \beta_{1}\left(\frac{\alpha_{s}}{2
\pi}\right)^{3}+{\mathcal O}\left(\alpha_{s}^{4}\right)
\label{betafun1}
\end{equation}
with $$\beta_{0}=\!\frac{11}{3}C_{A}\!-\!\frac{4}{3}T_{R}N_{f},$$
$$\beta_{1}=\!
\frac{34}{3}C_{A}^{2}\!-\!4\left(\frac{5}{3}C_{A}\!+\!C_{F}\right)T_{R}N_{f}$$
 and  $C_F=\frac{\left(N_{c}^{2}-1
\right)}{2 N_{c}}$, $C_A=N_c$ and $T_R=1/2$ \cite{Gross:1973id,Politzer:1973fx,Caswell:1974gg,Jones:1974mm,Egorian:1978zx}). \\
 Being this a first order differential equation we need an initial value of $\alpha_s$
at a given energy scale. This value is determined
phenomenologically. In QCD the number of colors $N_c$ is set to 3,
while the $N_f$ , i.e. the number of active flavors, varies across
the quark mass thresholds. In this work we determine the evolution
of the strong coupling keeping the number of colors fixed and
varying the number of flavors within the perturbative regime of
the conformal window.

\subsection{Two-loop results}
\begin{figure}[htb]
\centering
\includegraphics[width=0.40\textwidth]{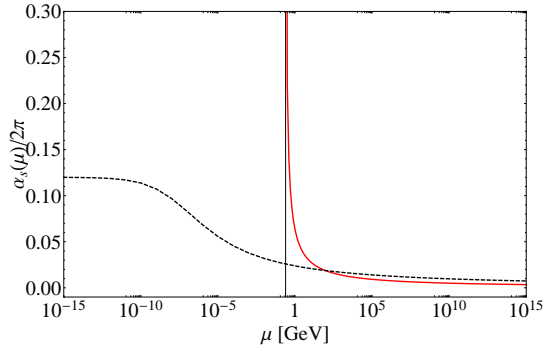}
\caption{The strong running coupling $\alpha_s(\mu)$ for $N_f=12$
(Black Dashed) and for $N_f=5$ (Solid Red). } \label{Lambert}
\end{figure}

In order to determine the solution for the strong coupling
$\alpha_s$ evolution we first introduce the following notation:
$x(\mu)\equiv \frac{\alpha_s(\mu)}{2 \pi}$,
$t=\log(\mu^2/\mu_0^2)$, $B=\frac{1}{2}\beta_0$ and
$C=\frac{1}{2}\frac{\beta_1}{\beta_0}$, $x^*\equiv -\frac{1}{C}$.
The truncated NNLO approximation of the Eq. \ref{betafun1} leads
to the differential equation:
\begin{equation}
\frac{dx}{dt}=-B x^2(1+C x) \label{lambert1}
\end{equation}
An implicit solution of Eq. \ref{lambert1} is given by the Lambert
$W(z)$ function:
\begin{equation}
W e^W = z \label{W}
\end{equation}
with: $ W=\left(\frac{x^*}{ x}-1\right)$. The general solution for
the coupling is:
\begin{eqnarray}
x &=& \frac{x^*}{1+W} , \\
 z &=& e^{\frac{x^*}{x_0}-1} \left(\frac{x^*}{x_0}-1 \right) \left( \frac{\mu^2}{\mu_0^2}
\right)^{x^* B}. \label{xz}
\end{eqnarray}
We will discuss here the solutions to the Eq. \ref{lambert1}
with respect to the particular initial phenomenological value
$x_0\equiv \alpha_s(M_{Z_0}) /(2\pi)= 0.01876 \pm 0.00016$ given
by the coupling determined at the $Z_0$ mass
scale~\cite{ParticleDataGroup:2020ssz}. In the range $N_f<\frac{34
N_c^3}{13 N_c^2-3}$ and $N_f>\frac{11}{2}N_c$ we have that the
solution is given by the $W_{-1}$ branch, while for $\frac{34
N_c^3}{13 N_c^2-3}< N_f < \frac{11}{2}N_c  $ the solution for the
strong coupling is given by the $W_{0}$ branch. By introducing the
phenomenological value $x_0$, we define a restricted range for the
IR fixed point discussed by Banks and Zaks~\cite{bankszaks}. Given
the value $\bar{N}_f =x^{*-1}(x_0) = 15.222 \pm 0.009$, we have
that in the range $\frac{34 N_c^3}{13 N_c^2-3}< N_f<\bar{N}_f$ the
$\beta$-function has both a UV and an IR fixed point, while for
$N_f> \bar{N}_f$ we no longer have the correct UV behavior. Thus
the actual physical range of a conformal window for pQCD is given
by $\frac{34 N_c^3}{13 N_c^2-3}< N_f<\bar{N}_f$.  The behavior of
the coupling is shown in Fig. \ref{Lambert}. In the IR region the
strong coupling approaches the IR finite limit, $x^*,$ in the case
of values of $N_f$ within the conformal window (e.g. Black Dashed
curve of Fig. \ref{Lambert}), while it diverges at
\begin{equation} \Lambda= \mu_0 \left(1+ \frac{|x^*|}{x_0}
\right)^{\frac{1}{2 B |x^*|}} e^{-\frac{1}{2 B x_0}}
\label{landaupole}\end{equation} outside the conformal window
given the solution for the coupling with $W_{-1}$ (e.g. Solid Red
curve of Fig. \ref{Lambert}). The solution of the NNLO equation
for the case $B>0, C>0$, i.e. $N_f<\frac{34 N_c^3}{13 N_c^2-3}$ ,
can also be given using the standard QCD scale parameter $\Lambda$
of Eq. \ref{landaupole},
\begin{eqnarray}
x &=& \frac{x^*}{1+W_{-1}} , \\
 z &=& -\frac{1}{e}  \left( \frac{\mu^2}{\Lambda^2}
\right)^{x^* B}. \label{xz}
\end{eqnarray}
This solution can be related to the one obtained in Ref.
\cite{Gardi:1998qr} by a redefinition of the $\Lambda$ scale.
 We underline that the presence of a
Landau "ghost" pole in the strong coupling is only an effect of
the breaking of the perturbative regime, including
non-perturbative contributions, or using non-perturbative QCD, a
finite limit is expected at any $N_f$~\cite{Deur:2016tte}. Both
solutions have the correct UV asymptotic free behavior. In
particular, for the case $\bar{N}_f<N_f<\frac{11}{2}N_c$, we have
a negative $z$, a negative $C$ and a multi-valued solution, one
real and the other imaginary, actually only one (the real) is
acceptable given the initial conditions, but this solution is not
asymptotically free. Thus we restrict our analysis to the range
$N_f<\bar{N}_f$ where we have the correct UV behavior. In general
IR and UV fixed points of the $\beta$-function can also be
determined at different values of the number of colors $N_c$
(different gauge group $SU(N)$) and $N_f$ extending this analysis
also to other gauge theories~\cite{Ryttov:2017khg}.

\subsection{Thrust at NNLO}

The thrust ($T$) variable is defined as
\begin{eqnarray}
T=\frac{\max\limits_{\vec{n}}\sum\limits_{i}|\vec{p}_i\cdot\vec{n}|}{\sum\limits_{i}|\vec{p}_i|},
\end{eqnarray}
where the sum runs over all particles in the hadronic final state,
and the $\vec{p}_i$  denotes the three-momentum of particle $i$.
The unit vector $\vec{n}$ is varied to maximize thrust $T$, and
the corresponding $\vec{n}$ is called the thrust axis and denoted
by $\vec{n}_T$. It is often used the variable $(1-T)$, which for
the LO of the 3 jet production is restricted to the range
$(0<1-T<1/3)$. We have a back-to-back or a spherically
symmetric event respectively at $T=1$ and at $T=2/3$ respectively.\\
 In general a normalized IR safe single variable observable, such as
the thrust distribution for the $e^+ e^-\rightarrow 3jets$
\cite{DelDuca:2016ily,DelDuca:2016csb}, is the sum of pQCD
contributions calculated up to NNLO at the initial renormalization
scale $\mu_0=\sqrt{s}=M_{Z_0}$:
\begin{eqnarray}
\frac{1}{\sigma_{tot}} \! \frac{O d \sigma(\mu_{0})}{d O}\! & = &
\left\{ x_0 \cdot \frac{ O d \bar{A}_{\mathit{O}}(\mu_0)}{d O} +
x_0^2 \cdot \frac{ O d \bar{B}_{\mathit{O}}(\mu_0)}{d O} \right. \nonumber  \\
 & & + \left. x_0^{3} \cdot \frac{O d\bar{C}_{\mathit{O}}(\mu_0)}{d
O}+ {\cal O}(\alpha_{s}^4) \right\},
 \label{observable1}
\end{eqnarray}
where $O$ is the selected Event Shape variable, $\sigma$ the cross section of the process,\\
$$\sigma_{tot}=\sigma_{0} \left( 1+x_0
A_{t o t}+ x_0^{2} B_{t o t}+ {\cal
O}\left(\alpha_{s}^{3}\right)\right)$$ \\ is the total hadronic
cross section and $\bar{A}_O, \bar{B}_O, \bar{C}_O$ are
respectively the normalized LO, NLO and NNLO coefficients:
\begin{eqnarray}
\bar{A}_{O} &=&A_{O} \nonumber \\
\bar{B}_{O} &=&B_{O}-A_{t o t} A_{O} \\
\bar{C}_{O} &=&C_{O}-A_{t o t} B_{O}-\left(B_{t o t}-A_{t o
t}^{2}\right) A_{O}. \nonumber
\end{eqnarray}
where $A_O, B_O, C_O$ are the coefficients normalized to the tree
level cross section $\sigma_0$ calculated by MonteCarlo (see e.g.
EERAD and Event2 codes~\cite{Gehrmann-DeRidder:2007nzq,
GehrmannDeRidder:2007hr, Ridder:2014wza, Weinzierl:2008iv,
Weinzierl:2009ms}) and $A_{\mathit{tot}}, B_{\mathit{tot}}$ are:
\begin{eqnarray}
A_{\mathit{tot}} &= & \frac{3}{2} C_F ; \nonumber \\
B_{\mathit{tot}} &= & \frac{C_F}{4}N_c +\frac{3}{4}C_F
\frac{\beta_0}{2} (11-8\zeta(3)) -\frac{3}{8} C_F^2.
 \label{norm}
\end{eqnarray}
where $\zeta$ is the Riemann zeta function.


 According to the PMC$_\infty$ (for an introduction on
the PMC$_\infty$ see Ref. \cite{DiGiustino:2020fbk})
Eq.\ref{observable1} becomes:
\begin{equation}
\frac{1}{\sigma_{tot}} \! \frac{O d
\sigma(\mu_I,\tilde{\mu}_{II},\mu_{0})}{d O}=
\left\{\overline{\sigma}_{I}+\overline{\sigma}_{II}+\overline{\sigma}_{III}+
{\cal O}(\alpha_{s}^4) \right\},
 \label{observable3}
\end{equation}
where the $\overline{\sigma}_{N}$ are normalized subsets that are
given by:
\begin{eqnarray}
\overline{\sigma}_{I} &=& A_{\mathit{Conf}} \cdot x_I \nonumber  \\
\overline{\sigma}_{II} &=& \left( B_{\mathit{Conf}}+\eta
A_{\mathit{tot}} A_{\mathit{Conf}} \right)\cdot x_{II}^2
 - \eta A_{\text{tot}} A_{\mathit{Conf}} \cdot x_0^2 \nonumber \\
 & & -A_{\text{tot}} A_{\mathit{Conf}}\cdot x_0 x_I  \nonumber \\
\overline{\sigma}_{III} &=&\!\! \left( C_{\mathit{Conf}} \!-\!
A_{\text {tot}} \!B_{\mathit{Conf}}\!-\!(B_{\text
{tot}}-A_{\text {tot}}^{2}) A_{\mathit{Conf}}\right) \cdot x_0^3 ,\nonumber \\
\label{normalizedcoeff}
\end{eqnarray}
$A_{\mathit{Conf}}, B_{\mathit{Conf}}, C_{\mathit{Conf}}$ are the
scale invariant conformal coefficients (i.e. the coefficients of
each perturbative order not depending on the scale $\mu_R$) while
$x_I,x_{II},x_0$ are the couplings determined at the
$\mu_I,\tilde{\mu}_{II},\mu_0$ scales respectively.  The
PMC$_\infty$ scales, $\mu_N$ , are given by:

\begin{eqnarray}
\large{\mu}_{\mathit{I}} & = & \sqrt{s} \cdot e^{f_{sc}-\frac{1}{2} B_{\beta_0}},\hspace{1.9cm}{ \scriptstyle (1-T)<0.33}  \nonumber \\
\large{\tilde{\mu}}_{\mathit{II}} & =& \left\{ \begin{array}{lr}
\sqrt{s} \cdot e^{f_{sc}-\frac{1}{2} C_{\beta_0} \cdot
\frac{B_{\mathit{Conf}}}{B_{\mathit{Conf}}+\eta \cdot
A_{\mathit{tot}} A_{\mathit{Conf}} }},\\ \hspace{3.9cm} {\scriptstyle (1-T)<0.33}  \\
  \sqrt{s}\cdot e^{f_{sc}-\frac{1}{2} C_{\beta_0}},\\ \hspace{3.9cm}{ \scriptstyle (1-T)>0.33}
  \label{icfscale}\\
 \end{array} \right.
 \label{PMC12}
\end{eqnarray}

and $\mu_0=M_{Z_0}$. The renormalization scheme factor for the QCD
results is set to $f_{sc}\equiv 0$. The coefficients
$B_{\beta_0},C_{\beta_0}$ are the coefficients related to the
$\beta_0$-terms of the NLO and NNLO perturbative order of the
thrust distribution respectively. They are determined from the
calculated $A_O, B_O, C_O$ coefficients by using the iCF ({\it the
intrinsic conformality} \cite{DiGiustino:2020fbk}).

The $\eta$ parameter is a regularization term in order to cancel
the singularities of the NLO scale, $\mu_{II}$, in the range
$(1-T)<0.33$, depending on non-matching zeroes between numerator
and denominator in the $C_{\beta_0}$. In general this term is not
mandatory for applying the PMC$_{\infty}$, it is necessary only in
case one is interested to apply the method all over the entire
range covered by the thrust, or any other observable. Its value
has been determined to $\eta=3.51$ for the thrust distribution and
it introduces no bias effects up to the accuracy of the
calculations and the related errors are totally negligible up to
this stage.

\section{The thrust distribution according to $N_f$}

\begin{figure}[htb]
 \centering
\hspace*{-0.5cm}
\includegraphics[width=0.50\textwidth]{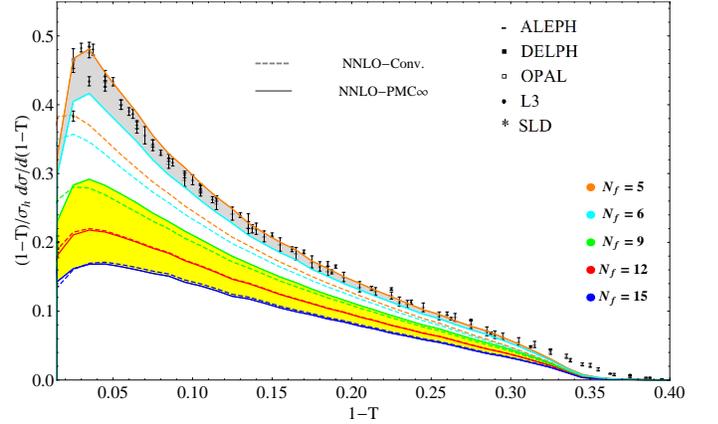}
\caption{Thrust distributions for different values of $N_f$, using
the PMC$_\infty$ (Solid line) and the CSS (Dashed line). The
Yellow shaded area is the results for the values of $N_f$ taken in
the conformal window. The experimental data points are taken from
the ALEPH, DELPHI,OPAL, L3, SLD experiments
\cite{aleph,delphi,opal,l3,sld}.} \label{thrust}
\end{figure}

Results for the thrust distribution calculated using the  NNLO
solution for the coupling $\alpha_s(\mu)$, at different values of
the number of flavors, $N_f$,  is shown in Fig. \ref{thrust}.

\begin{figure}[htb]
\centering
\includegraphics[width=0.4\textwidth]{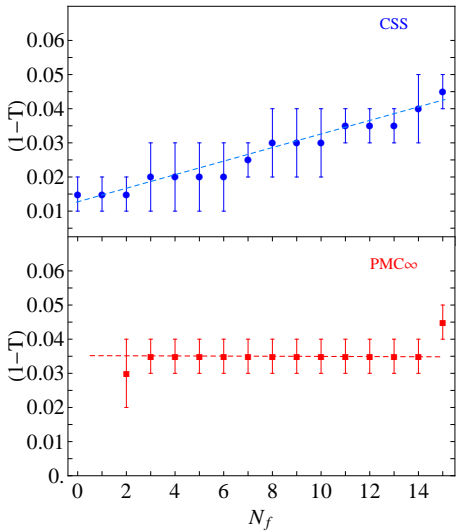}
\caption{Comparison of the position of the peak for the thrust
distribution using the CSS and the PMC$_\infty$ vs the number of
flavors, $N_f$. Dashed Lines indicate the particular trend in each
graph.} \label{peaks}
\end{figure}

 A direct comparison between PMC$_\infty$ (Solid line) and CSS
(Dashed line) is shown at different values of the number of
flavors. We notice that, despite the phase transition (i.e. the
transition from an infrared finite coupling to an infrared
diverging coupling), the curves given by the PMC$_\infty$ at
different $N_f$, preserve with continuity the same characteristics
of the conformal distribution setting $N_f$ out of the conformal
window of pQCD. In fact, the position of the peak of the thrust
distribution is well preserved varying the $N_f$ in and out of the
conformal window using the PMC$_\infty$, while there is constant
shift towards lower values using the CSS. These trends are shown
in Fig. \ref{peaks}. We notice that in the central range,
$2<N_f<15$, the position of the peak is exactly preserved using
the PMC$_\infty$ and overlaps with the position of the peak shown
by the experimental data. According to our analysis for the case
PMC$_\infty$, in the range, $N_f<2$ the number of bins is not
enough to resolve the peak, though the behavior of the curve is
consistent with the presence of a peak in the same position, while
for $N_f \rightarrow 0$ the peak is no longer visible.
Theoretical uncertainties on the position of the peak have been
calculated using standard criteria, i.e. varying the remaining
initial scale value in the range $M_{Z_0}/2 \leq \mu_0 \leq 2
M_{Z_0}$, and considering the lowest uncertainty given by the half
of the spacing between two adjacent bins.

\begin{figure}[htb]
\centering
\includegraphics[width=0.4\textwidth]{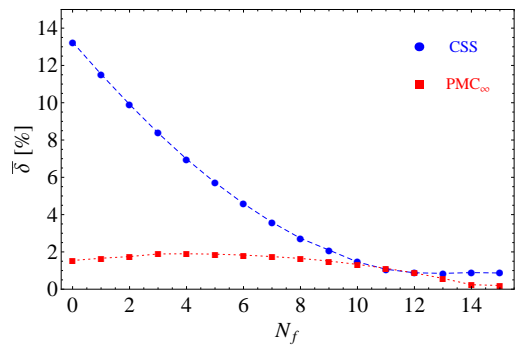}
\caption{Comparison of the average theoretical error,
$\bar{\delta}$, calculated using standard criteria in the range:
$0.005<(1-T)<0.4$, using the CSS and the PMC$_\infty$ for the
thrust distribution vs the number of flavors, $N_f$.} \label{err}
\end{figure}

 Using the definition given in Ref.~\cite{Gehrmann-DeRidder:2007nzq}
 of the parameter
 \begin{equation}
\delta=\frac{ \text{max}_{\mu}(\sigma(\mu))- \text{min}_{\mu}
(\sigma(\mu))}{2 \sigma(\mu=M_{Z_0})} , \label{delta}
\end{equation}
with the renormalization scale varying $\mu\in [M_{Z_0}/2; 2
M_{Z_0}]$, we have determined the average error, $\bar{\delta}$,
calculated in the interval $0.005<(1-T)<0.4$ of the thrust and
results for CSS and PMC$_\infty$ are shown in Fig. \ref{err}. We
notice that the PMC$_\infty$ in the perturbative and IR conformal
window, i.e. $12<N_f<\bar{N}_f$, which is the region where
$\alpha_s(\mu)<1$ in the whole range of the renormalization scale
values, from $0$ up to $\infty$, the average error given by
PMC$_\infty$ tends to zero ($\sim 0.23-0.26\%$) while the error
given by the CSS tends to remain constant ($0.85-0.89\%$). The
comparison of the two methods shows that, out of the conformal
window, $N_f<\frac{34 N_c^3}{13 N_c^2-3}$, the PMC$_\infty$ leads
to a higher precision.

\section{The thrust distribution in the Abelian limit $N_c\rightarrow 0$}

\begin{figure}[htb]
\centering
\includegraphics[width=0.40\textwidth]{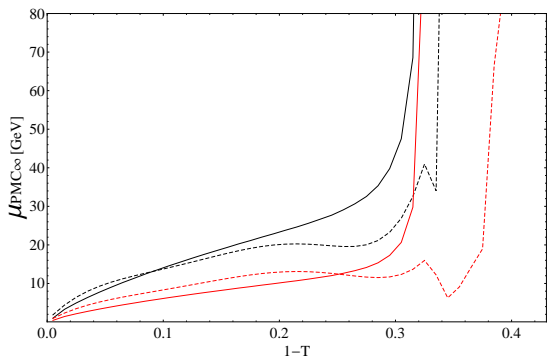}
\caption{PMC$_\infty$ scales for the thrust distribution: LO-QCD
scale (Solid Red); LO-QED scale (Solid Black);NLO-QCD scale
(Dashed Red); NLO-QED scale (Dashed Black).} \label{scales}
\end{figure}

We obtain the QED thrust distribution performing the
$N_c\rightarrow 0$ limit of the QCD thrust at NNLO according to
\cite{Brodsky:1997jk,Kataev:2015yha}. In the zero number of colors
limit the gauge group color factors are fixed by $N_A=1, C_F=1,
T_R=1, C_A=0,N_c=0, N_f=N_l$, where $N_l$ is the number of active
leptons, while the $\beta$-terms and the coupling rescale as
$\beta_n/C_F^{n+1}$ and $\alpha_s \cdot C_F$ respectively. In
particular $\beta_0=-\frac{4}{3}N_l$ and $\beta_1=-4 N_l$ using
the normalization of Eq. \ref{betafun1}. According to these
rescaling of the color factors we have determined the QED thrust
and the QED PMC$_\infty$ scales. For the QED coupling , we have
used the analytic formula for the effective fine structure
constant in the $\overline{\text{MS}}$-scheme:
\begin{equation}
{\alpha(Q^2)} = {\alpha \over  { \left(1 -\Re e
\Pi^{\overline{\text{MS}}} (Q^2)\right)}},
\end{equation}
with $\alpha^{-1}\equiv \alpha(0)^{-1}= 137.036$ and the vacuum
polarization function ($\Pi$) calculated perturbatively at two
loops including contributions from leptons, quarks and $W$ boson.
The QED PMC$_\infty$ scales have the same form of Eq.
\ref{icfscale} with the factor for the
$\overline{\text{MS}}$-scheme set to $f_{sc}\equiv 5/6$ and the
$\eta$ regularization parameter introduced to cancel singularities
in the NLO PMC$_\infty$ scale $\mu_{II}$ in the $N_c \rightarrow
0$ limit tends to the same QCD value, $ \eta=3.51 $. A direct
comparison between QED and QCD PMC$_\infty$ scales is shown in
Fig. \ref{scales}.

\begin{figure}[htb]
\centering \vspace{1cm}
\includegraphics[width=0.40\textwidth]{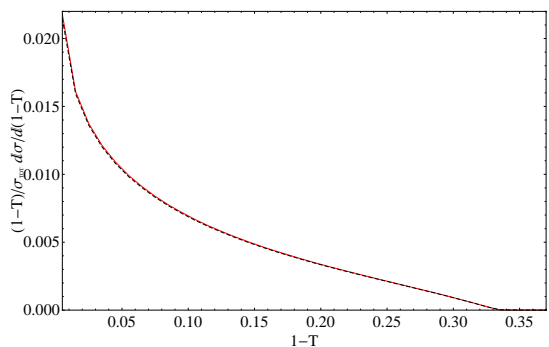}
\caption{Thrust distributions in the QED limit at NNLO using the
PMC$_\infty$ (Solid Red) and the CSS (Dashed Black).}
\label{qedthrust}
\end{figure}

We notice that in the QED limit the PMC$_\infty$ scales have
analogous dynamical behavior as those calculated in QCD,
differences arise mainly by the $\overline{\text{MS}}$ scheme
factor reabsorption, the effects of the $N_c$ number of colors at
NLO are very small. Thus we notice that perfect consistency is
shown from QCD to QED using the PMC$_\infty$ method. The
normalized QED thrust distribution is shown in Fig.
\ref{qedthrust}. We notice that the curve is peaked at the origin,
$T=1$, which suggests that the three jet event in QED occurs with
a rather back-to-back symmetry. Results for the CSS and the
PMC$_\infty$ methods in QED are of the order of $O(\alpha)$ and
given the good convergence of the theory the results for the two
methods show very small differences.

\section{Conclusions}

We have investigated, for the first time, the thrust distribution
in the conformal window of pQCD. Assuming, for phenomenological
reasons, the physical value of the strong coupling to be the one
at the $Z_0$ mass scale it restricts the conformal window range,
at two loops, to be within $\frac{34 N_c^3}{13 N_c^2-3}<
N_f<\bar{N}_f$ with $ \bar{N}_f\simeq 15.22$. The closer $N_f$ to
the higher value the more perturbative and conformal the system
is. In this region, we have shown that the PMC$_\infty$ leads to a
higher precision with a theoretical error which tends to zero.
Besides results for the thrust distribution in the conformal
window have similar shapes to those of the physical values of
$N_f$ and the position of the peak is preserved when one applies
the PMC$_\infty$ method. Comparison with the experimental data
indicates also that PMC$_\infty$ agrees with the expected number
of flavors. A good fit with experimental data is shown by the
PMC$_\infty$ results for the range $5<N_f<6$, which agrees with
the active number of flavors of the Standard Model. Outside the
pQCD conformal window the PMC$_\infty$ leads to a higher precision
with respect to the CSS. In addition, calculations for the QED
thrust reveal a perfect consistency of the PMC$_\infty$ with QED
when taking the QED limit of QCD for both the PMC$_\infty$ scale
and for the regularization $\eta$ parameter which tends to the
same QCD value.

{\bf Acknowledgements}: We thank Stanley J. Brodsky for his
valuable comments and for useful discussions. We thank Einan Gardi
and Philip G. Ratcliffe for helpful discussions.


\end{document}